\documentstyle[aps]{revtex}

\def\bm#1{\mbox{\boldmath$#1$}}
\newcommand{\be}{\begin{equation}}
\newcommand{\ee}{\end{equation}}
\newcommand{\bc}{\begin{center}}
\newcommand{\ec}{\end{center}}
\newcommand{\ba}{\begin{array}}
\newcommand{\ea}{\end{array}}
\newcommand{\bt}{\begin{tabular}}
\newcommand{\et}{\end{tabular}}

\begin{document}
\draft

\title{\large {\bf A NUMBER PROJECTED  MODEL WITH
GENERALIZED   PAIRING INTERACTION
\\
}}

\author{ {\bf W. Satu{\l}a}$^{1-4}$ and
{\bf R. Wyss}$^{2,5}$ \\
}

\address{
$^1$ Institute of Theoretical Physics, Warsaw University, ul. Ho\.za 69,
PL-00 681, Warsaw, Poland\\
$^2$ Royal Institute of Technology, Physics Department Frescati,
Frescativ\"agen 24, S-104 05 Stockholm, Sweden \\
$^3$ Joint Institute for Heavy Ion Research, Oak Ridge National Laboratory \\
P.O. Box 2008, Oak Ridge, TN 37831, U.S.A.\\
$^4$ Department of Physics, University of Tennessee, Knoxville, TN 37996,
U.S.A.  \\
$^5$ Department of Technology, Kalmar University, \\
Box 905, 391 29 Kalmar, Sweden}

\date{\today}

\maketitle

\begin{abstract}
A mean-field model with a generalized pairing
interaction that accounts for neutron-proton pairing
is presented. Both the BCS as well as number-projected
solutions of the model are presented. For the latter case
the Lipkin-Nogami projection technique was
extended to encompass the case of non-separable proton-neutron systems.
The influence of the projection on various pairing phases is
discussed. In particular, it is shown that number-projection allows
for mixing of different pairing phases but, simultanously, acts
destructively on the proton-neutron correlations.
The basic implications of proton-neutron pairing correlations
on nuclear masses are discussed. It is shown that these correlations
may provide a natural microscopic explanation of the Wigner energy
lacking in mean-field models.
A possible phase
transition from isovector to isoscalar pairing condensate at high
angular momenta is also discussed. In particular predictions
for the dynamical moments of inertia for the  superdeformed band
in $^{88}$Ru are given.

\end{abstract}


\newpage

\section{Introduction}\label{s1}

  The strongest effects associated with neutron-proton ({\it np\/}) pairing
are expected in $N\approx Z$ nuclei where valence protons and neutrons occupy
the same shell-model orbits. The basic properties of the
$np$-interaction are known
from the studies of simple configurations of near closed
shell nuclei~\cite{[Ana71],[Sch71]}. The isovector (T=1) interaction
is dominated by the J=0$^+$ channel but the isoscalar (T=0)
$np$-interaction is almost equally attractive in the
J=1$^+$ and stretched J=(2$j$)$^+$ channels.
The T=0 interaction is on the average stronger than the T=1 force.
It may, therefore, lead to the appearance of a static T=0
$np$-pairing condensate,
particularly in heavier $N\approx Z$ nuclei
where the large valence space allows for the creation of  many {\it np}-pairs.
However, it is not obvious
whether these correlations are coherent enough to create
this new type of collective mode nor what are the main
building blocks or specific experimental fingerprints of such a
condensate.

Theoretically, $np$-pairing is  a challenging subject.
It offers new opportunities to probe specific parts
of the effective nucleon-nucleon interaction.
The generalization of BCS (or HFB) techniques to incorporate and allow
for unconstrained interplay of T=0 and T=1 pairs on equal footing is
by itself non-trivial. Though the
first steps to generalize the BCS theory as well as
the first applications were done
already in the sixties
\cite{[Gos64],[Gos65],[Che67],[Goo68],[Bar69],[Goo70],[Wol70],[Wol71],[Goo72]}
(for review of the early efforts see~\cite{[Goo79]})
only recently the first symmetry unconstrained, self-consistent
mean-field calculations have been performed\cite{[Ter98]}.
Extensions beyond mean-field, restoring either
rigorously or approximately number symmetry and/or isospin symmetry are
scarce.

The renaissance in the interest for $np$-pairing
can be traced back to the fast progress in detection techniques and
radioactive ion beam (RIB) programs. First
experiments with RIB's are soon to come and are targeted on heavy
proton-rich nuclei in particular on $N\approx Z$ nuclei.
They are expected to provide important clues
resolving the above mentioned, long standing difficulties in understanding
$np$-pairing. The observables to look for are obviously those which are
expected to be strongly modified by a static $np$-pair condensate
like deuteron-transfer probability~\cite{[Fro70],[Fro71]},
$\beta $ and Gamow-Teller decay rates~\cite{[Sch96],[Eng97]} or ground-state
and high-spin
properties~\cite{[Nic78],[Mul82],[Sat97],[Goo98],[Goo99]}.

So far no clear, systematic experimental signature of
the $np$-pairing condensate is known. There are, however, some indirect
indications, for example, in recent spectroscopic
data in $^{72}_{36}$Kr$_{36}$~\cite{[Ang97]} and
$^{74}_{37}$Rb$_{37}$~\cite{[Rud96]}.
In the ground state of $^{74}_{37}$Rb$_{37}$, with {\bf T}=1,{\bf T}$_z$=0,
the
$\gamma-$ray energies of the collective $4^+\rightarrow 2^+\rightarrow 0^+$
transitions  appear to be similar (isobaric analogues) to
$^{74}_{36}$Kr$_{38}$, the {\bf T}=1,{\bf T}$_z$=1 nucleus in spite of the
expected increase in the dynamical moment of inertia due to blocking of
the like-particle superfluidity.\footnote{Throughout the paper, the bold-faced
symbols {\bf T} and {\bf T}$_z$=(N-Z)/2 would refer to the total nuclear
isospin and its z-component, respectively. The T and T$_z$ are reserved to
distinguish between various two-body interaction channels.}
This phenomenon  has been interpreted as a manifestation of T=1 $np$-pairing
collectivity~\cite{[Rud96]}.
At higher spins a transition from T=1 to T=0 band has also been observed.
Calculations seem to confirm the T=1 $np$-collectivity at low-spins and
predict an increasing role of the aligned T=0 pairs at higher spins, see
Refs.~\cite{[Dea97],[Pet99],[Fra99]}.
In $^{72}_{36}$Kr$_{36}$ a rather unexpected
delay of the first crossing frequency has been measured~\cite{[Ang97]}.
It again may have possible links to $np$-pairing, see discussion
in~\cite{[Dea97],[Fra99],[Pet99],[Fra99a]},
although more conventional
explanations involving shape vibrations cannot be ruled out.

The strongest evidence for the enhancement of $np$-pairing effects seems
to come from binding energies. The well known slope discontinuity of the
isobaric mass parabola at
$N\approx Z$, see review~\cite{[Zel96]} and refs. therein,
indicates an additional binding energy (Wigner energy) in $N\approx Z$ nuclei.
The  Wigner energy is predominantly due to the T=0
interaction~\cite{[Bre90],[Sat97a]}.
However, the mechanism responsible for the extra binding energy
seems to be rather complex when expressed in terms of $np$-pairs
of given J,T~\cite{[Sat97a]}. It cannot be solely
explained in terms of J=1,T=0
$np$-pairs, at least not for $sd$ or $pf$ shell nuclei.
A connection of the Wigner energy and the T=0 $np$-pairing
condensate was suggested in our Letter~\cite{[Sat97]} based
on deformed mean-field calculations with a schematic pairing interaction.
Mass measurements of more heavy $N\approx Z$ nuclei are needed to
shed more light on this issue.

The aim of this paper is to further investigate basic features of $np$-pairing.
The paper supplements the above mentioned
letter~\cite{[Sat97]} explaining in more detail certain
technical aspects of our model but also provides new numerical
and analytical results. The paper is organized as follows:
In  Sect.~\ref{s2} we introduce the
basic concepts concerning the Bogolyubov transformation and
the self-consistent symmetries (SCS) used here to simplify the calculations.
Details concerning the model hamiltonian and implications of SCS
on the structure
and interpretation of the model can be found in Section~\ref{s3}.
Section~\ref{s4} presents the method used to restore approximatively
the particle-number symmetry which is an extension
of the so called Lipkin-Nogami technique for the case of a non-separable
proton-neutron system. The ideas presented in this section
are independent on the kind of two-body interaction used in the
calculations.
The results of numerical calculations, discussion and conclusions are
given in Sections~\ref{s5} and~\ref{s6}, respectively.

\section{The Bogolyubov transformation and self-consistent
symmetries}\label{s2}

The starting point of our considerations are the
eigenstates of a {\it deformed} phenomenological
single-particle potential.
The basis states can be divided into two groups with respect to
the signature symmetry ($\hat R_x = e^{-i\pi \hat{j_x}}$) quantum
number $r=-i(+i)$ which are later
labeled as {\bm \alpha}($\tilde {\bm {\alpha}}$), respectively.
Two different types of nucleonic pairs can therefore be formed, namely
${\bm {\alpha}}\tilde{\bm {\beta}}$ and {\bm {\alpha\beta}} pairs.
A generalized BCS (gBCS) theory has to account for a
scattering of these two types of nucleonic pairs  simultaneously.
In fact, this work is restricted to pairing
in the same and signature reversed states i.e. for
${\bm {\alpha}}\tilde{\bm {\alpha}}$ and {\bm {\alpha\alpha}}
modes only, see also~\cite{[Goo72]}.

The signature symmetry cannot be used as a
self-consistent symmetry (SCS) in gBCS calculations. Indeed, in such a case
the pairing tensor, ${\bm {\kappa =V}}^*{\bm U}^T$, connects only states of
opposite signature~\cite{[Goo74]}.
Consequently, the {\bm {\alpha\alpha}}  $np$-pairing cannot be activated,
see also~\cite{[Nic78],[Mul82]}.
Therefore, to take into account simultaneously
${\bm {\alpha}}\tilde{\bm {\alpha}}$
and  {\bm {\alpha\alpha}} pairing one needs to extend the Bogolyubov
transformation. The most general Bogolyubov transformation can be written
as:
\be\label{eq3}
\hat\alpha^\dagger_k = \displaystyle\sum_{\alpha >0}
(U_{\alpha k} a^\dagger_{\alpha}
+V_{\tilde\alpha k} a_{\tilde\alpha}+ U_{\tilde\alpha  k}
a^\dagger_{\tilde \alpha}
+V_{\alpha k} a_{\alpha})
 \ee
where $\alpha (\tilde\alpha)$ denote {\it single particle}
states (including isospin indices) of signature $r=-i(+i)$ respectively,
while $k$ denotes quasiparticles. As discussed above,
by superimposing any SCS one always excludes certain
interaction channels in the mean-field approximation.
Nevertheless, the use of SCS appear many times inherent to
the nature of the physical problem and of course,
make the theory more transparent and easier
to handle numerically. Therefore, prior to construct a general
theory involving the transformation (\ref{eq3}) we
simplify the problem by superimposing (beyond parity) the so called
{\it antilinear simplex symmetry\/},
$\hat S_{x}^{T}=\hat P\hat T\hat R_{z}$,
as SCS, see~\cite{[Goo74],[Nic78]}.
One should bear in mind  that due to the antilinearity of $\hat S_{x}^{T}$
the transformation properties of creation and destruction operators
with respect to $\hat S_{x}^{T}$ will depend on the phases of
the basis states, i.e. no new quantum number can by related
directly to this symmetry.

When superimposing $\hat S_{x}^{T}$ as the SCS it
is rather convenient to choose the phase convention
in such a way that the basis states will have
exactly
the same transformation properties with respect to both $\hat R_x$
and $\hat S_{x}^{T}$ and:
\be\label{eq4}
 \ba{ccc}  &
  \hat S_{x}^{T}
   \left(
     \ba{c}
        a^\dagger_\alpha \\ a^\dagger_{\tilde\alpha}
     \ea
   \right)
 (\hat S_{x}^{T})^{-1}=
   i  \left(
      \ba{r}
        - a^\dagger_\alpha \\ a^\dagger_{\tilde\alpha}
      \ea
      \right)
 &  \ea \ee
Let us divide our quasiparticle states (\ref{eq3}) into two families
denoted as ${k}$ and ${\tilde k}$, respectively:
\be\label{eq3x}\ba{ccc}
\hat\alpha^\dagger_k&=&\displaystyle\sum_{\alpha >0}
(U_{\alpha k} a^\dagger_{\alpha}
+V_{\tilde\alpha k} a_{\tilde\alpha}+ U_{\tilde\alpha  k}
a^\dagger_{\tilde \alpha}
+V_{\alpha k} a_{\alpha}) \\
\hat\alpha^\dagger_{\tilde k}&=&\displaystyle\sum_{\alpha >0}
(U_{\tilde\alpha \tilde k}
\hat a^\dagger_{\tilde\alpha} +V_{\alpha \tilde k} \hat a_{\alpha}
+ U_{\alpha \tilde k}
\hat a^\dagger_{\alpha} +V_{\tilde\alpha \tilde k} \hat a_{\tilde\alpha}).
\ea \ee
Enforcing $\hat S_{x}^{T}$ symmetry as
SCS requires that the quasiparticle
operators of eq.~(\ref{eq3x}) have the same
transformation properties with respect to $\hat S_{x}^{T}$
as the single particle operators (\ref{eq4})~\cite{[Goo74]}.
This leads to the following restrictions for the coefficients
of the Bogolyubov transformation (\ref{eq3x}):
\be\label{eq6}
    \left\{ \ba{c}
         U_{\alpha k} = U^*_{\alpha k} \\
    U_{\tilde\alpha \tilde k} = U^*_{\tilde\alpha \tilde k} \\
         V_{\tilde\alpha k} = V^*_{\tilde\alpha k} \\
         V_{\alpha \tilde k} = V^*_{\alpha \tilde k} \\
            {\bm {Real}}
    \ea \right\}
         \quad \quad
    \left\{ \ba{c}
         U_{\tilde\alpha k} = -U^*_{\tilde\alpha k} \\
    U_{\alpha \tilde k} = -U^*_{\alpha \tilde k} \\
       V_{\tilde\alpha \tilde k} = -V^*_{\tilde\alpha \tilde k} \\
         V_{\alpha k} = -V^*_{\alpha k} \\
         {\bm {Imaginary}}
    \ea \right\}.
\ee
The formalism is complex but the
density matrix, ${\bm {\rho=V}}^* {\bm V}^T$, and  pairing tensor
${\bm {\kappa = V}}^*{\bm U}^T$ take the relatively simple structure with
real and imaginary blocks decoupled from each other:
\be\label{eq7}
{\bm \rho}= \left( \ba{cc}
             \mbox{{\bf Re}}(\rho_{\alpha\beta}) & 0 \\
             0 & \mbox{{\bf Re}}(\rho_{\tilde\alpha\tilde\beta})
      \ea \right) +
 i   \left( \ba{cc}
             0 & \mbox{{\bf Im}}(\rho_{\alpha\tilde\beta}) \\
             \mbox{{\bf Im}}(\rho_{\tilde\alpha\beta}) & 0
     \ea \right)
\ee
\be\label{eq8}
{\bm \kappa}= \left( \ba{cc}
             0 & \mbox{{\bf Re}}(\kappa_{\alpha\tilde\beta})  \\
             \mbox{{\bf Re}}(\kappa_{\tilde\alpha\beta}) & 0
      \ea \right) +
 i   \left( \ba{cc}
             \mbox{{\bf Im}}(\kappa_{\alpha\beta}) & 0\\
             0 & \mbox{{\bf Im}}(\kappa_{\tilde\alpha\tilde\beta})
     \ea \right)
\ee
Furthermore, the complex structure  of the single particle potential
${\bm \Gamma}$ and the pairing potential ${\bm \Delta}$
\be
\Gamma_{\alpha\beta}\equiv \displaystyle\sum_{\gamma\delta}
{\bar v}_{\alpha\gamma\beta\delta}\rho_{\delta\gamma}
\quad \mbox{and}\quad
\Delta_{\alpha\beta}\equiv {1\over 2}\displaystyle\sum_{\gamma\delta}
{\bar v}_{\alpha\beta\gamma\delta}\kappa_{\gamma\delta}
\ee
and consequently the gBCS equations are fully determined by
the {\bm \rho} and {\bm \kappa} matrices, respectively.

In this work we define the two-body $np$-pairing interaction
in terms of an extension of the standard seniority pairing
interaction. It is separable in the
particle-particle channel, ${\bar v}_{\alpha\beta\gamma\delta}\propto
g_{\alpha\beta}g^*_{\gamma\delta}$, with $g_{\alpha\beta}$ proportional
(up to a phase factor) to the overlap
$\langle\alpha_\tau | \beta_{\tau'}\rangle$ between the
single-particle wave functions.\footnote{Subscript $\tau$ in
 $\alpha_{\tau}$ is necessary to distinguish between
proton or neutron single-particle states.}
As already mentioned it takes essentially
${\bm {\alpha}}\tilde{\bm {\alpha}}$  and {\bm {\alpha\alpha}}
types of pairing, see Sect.~\ref{s3} for further detail.
To further visualize the physical implications of the $\hat S_{x}^{T}$
let us consider the limits of isospin ($N=Z$ case without Coulomb force) and
time reversal symmetry (non-rotating case).
Let us consider ${\bm {\alpha}}\overline{\bm {\alpha}}$ pairing which
in principle consists of both T=1 and T=0 components of the $np$-pairing.
By decomposing the pairing potential into the different isospin
components T,T$_z$ one finds:
\be
\Delta^{(1,0)}_{\alpha t_z,\overline{\alpha -t_z}} \propto
\kappa_{\alpha  1/2,\overline{\alpha -1/2}} +
\kappa_{\alpha -1/2,\overline{\alpha  1/2}}
\quad\mbox{and} \quad
\Delta^{(0,0)}_{\alpha t_z,\overline{\alpha -t_z}} \propto
\kappa_{\alpha  1/2,\overline{\alpha -1/2}} -
\kappa_{\alpha -1/2,\overline{\alpha  1/2}}
\ee
i.e. the T=1 and
T=0 components of $np$-pairing  depend on the combinations of
the same elements of the pairing tensor but with opposite sign.
This is due to the phase relation for
the Clebsh-Gordan coefficients which is (anti-)symmetric
with respect to the interchange of a proton and neutron for T=(0)1,
respectively. Time reversal symmetry further implies that~\cite{[Goo72]}:
\be
\kappa_{\alpha  1/2,\overline{\alpha -1/2}} =
-\kappa^*_{\overline{\alpha 1/2},\alpha -1/2} =
\kappa^*_{\alpha -1/2,\overline{\alpha 1/2}}
\ee
and therefore
\be
\Delta^{(1,0)}_{\alpha t_z,\overline{\alpha -t_z}} \propto
\mbox{{\bf Re}} (\kappa_{\alpha  1/2,\overline{\alpha -1/2}})
\quad\mbox{and} \quad
\Delta^{(0,0)}_{\alpha t_z,\overline{\alpha -t_z}} \propto
\mbox{{\bf Im}}(\kappa_{\alpha  1/2,\overline{\alpha -1/2}})
\ee
Consequently, with the pairing tensor of the form of  (\ref{eq8}),
the T=0 component of ${\bm {\alpha}}\overline{\bm {\alpha}}$
is ruled out through the $\hat S_{x}^{T}$
symmetry. Similar analysis shows that T=1 component of the
${\bm \alpha}{\bm \alpha}$ pairing also vanishes due to symmetry
reasons.

The latter is well justified due to the Pauli principle.
The lack of T=0 ${\bm {\alpha}}\overline{\bm {\alpha}}$ is a deficiency
of our model. However, for $\hbar\omega =0$ the
${\bm {\alpha}}\tilde{\bm {\alpha}}$ and
${\bm {\alpha}}{\bm {\alpha}}$ $np$-pairing phases are in many applications
indistinguishable due to time-reversal symmetry. Since our interaction
is essentially structureless, based on pair counting we expect our
results not to be sensitive to this restriction.
For $\hbar\omega\ne 0$, on the other hand, one wants to associate the
T=0, ${\bm {\alpha}}{\bm {\alpha}}$
(${\bm {\alpha}}\tilde{\bm {\alpha}}$) $np$-pairing
with the coupling to maximum (minimum) spin, respectively.
The retained component is expected to be dominant but only at
high spins. In order to probe the transition
from low to high spin regime one has to explore all possible T=0 pairs and
allow for an unconstrained interplay between the different pairing modes.
Note, however, that again due to the simplicity of our interaction
certain features of the transitional regime can be
simulated to some extent with an isospin broken hamiltonian. Indeed, the
missing T=0 ${\bm {\alpha}}\tilde{\bm {\alpha}}$ component is
expected to respond
to nuclear rotation in a similar way as
T=1 ${\bm {\alpha}}\tilde{\bm {\alpha}}$. It was shown explicitely
in Ref.~\cite{[Wys99]} for a single $j$-shell model.

In conclusion, in our model the $\bm{\alpha\tilde\alpha}$ pairing is
equivalent to T=1 and $\bm{\alpha\alpha}$ to T=0 and the isospin
notation will be used in the following.
This simple analysis reveals also the {\it important\/}
role the {\it self-consistent symmetries\/} can play
in theoretical description of $np$-pairing in the
mean-field theory, see also~\cite{[Wol71]}.

\section{The model Hamiltonian}\label{s3}

The
multipole-multipole expansion offers a rather good
approximation to the pairing energy
when neutrons and protons can be treated separably.
In the following we extend this idea to the case of
$np$-pairing by constructing a generalized
pairing force separable in the
particle-particle channel:
\be
\hat V_{pair} =
{1\over 4} G \displaystyle \sum \bar{v}_{\alpha\beta\gamma\delta}
\hat a^\dagger_\alpha \hat a^\dagger_\beta \hat a_\delta \hat a_\gamma
\equiv -{1\over 4} \displaystyle \sum_{\alpha\beta} g_{\alpha\beta}
 \hat a^\dagger_\alpha \hat a^\dagger_\beta
\circ \displaystyle \sum_{\gamma\delta} g_{\gamma\delta}^* \hat
a_\delta \hat a_\gamma
\ee
where $g_{\alpha\beta} \equiv \langle \alpha |
\hat G |\beta\rangle$ and  $\hat G$ is an auxiliary operator
generating the specific pairing mode with strength $G$.
The antisymmetry of the two-body matrix
element $\bar{v}_{\alpha\beta\gamma\delta}$  implies that
\be\label{anti}
\langle \alpha | \hat G | \beta \rangle = -
\langle \beta  | \hat G | \alpha \rangle \quad \forall
\quad { \alpha,\beta}.
\ee
and therefore, the generators $\hat G$ must be {\it antilinear\/} and
{\it antihermitian\/}. We assume here that
the correlation energy of the nucleonic pair is
proportional to the overlap $\langle \alpha_\tau | \beta_{\tau'}\rangle$
between single-particle states they occupy
(extended seniority-type pairing interaction). The $\hat G$ can then be
chosen, for example, as:
\be\label{gene}
\hat G_{\tau\tau}=\hat T, \quad
\hat G_{np}^{\alpha\tilde\alpha} =\hat G_{np}^{T=1} =
{1\over \sqrt{2}}\hat \tau_x \hat T, \quad
\hat G_{np}^{\alpha\alpha} = \hat G_{np}^{T=0}
={1\over \sqrt{2}}\hat \tau_y \hat S_{x}^{T}
\ee
for {\it pp\/}({\it nn\/}) pairing, ${\bm \alpha}\tilde{\bm \alpha}$
type of {\it np}-pairing and
${\bm {\alpha\alpha}}$ of {\it np}-pairing.
This choice is, however, not unique. In particular, it depends
on the choice of the relative phases between neutron and proton
states. The choice of phase convention induces strict transformation rules
for the isospin Pauli operators $\hat \tau_i,\, i=x,y,z$ with respect to
time reversal symmetry~\cite{[Dob96a]}.
Constructing the generators (\ref{gene}) we assumed the same phases for proton
and neutron states (~$|\alpha_\tau ,r ,\tau\rangle$ denotes the basis
state $\alpha$ of signature $r=\pm i$ and isospin $\tau= 1(-1)$
for neutrons(protons)):
\be
\hat T  |\alpha_\tau , r=\mp i, \tau \rangle  =
\mp |\alpha_\tau , r=\pm i, \tau \rangle
\ee
Our phase convention further implies that:
\be\ba{ccccc}
\hat T \hat \tau_x |\alpha_\tau ,r=\pm i,\tau\rangle  & = &
\hat T  |\alpha_\tau ,r=\pm i, -\tau\rangle & = &
\mp |{\alpha_\tau ,r=\mp i,  -\tau}\rangle \\
 \hat \tau_x \hat T |\alpha_\tau ,r=\pm i,\tau\rangle & = &
\mp \hat \tau_x  |\alpha_\tau ,r=\mp i, \tau\rangle  & = &
\mp |{\alpha_\tau ,r=\mp i,  -\tau}\rangle
\ea\ee
leading to $\hat T\hat \tau_x \hat T^{-1} = \hat \tau_x$. Similar
considerations for $\hat \tau_y$ and $\hat \tau_z$ operators give:
\be\label{phas}\ba{ccc}
\hat T \hat \tau_x \hat T^{-1} & =& + \hat \tau_x \\
\hat T \hat \tau_y \hat T^{-1} & =& - \hat \tau_y \\
\hat T \hat \tau_z \hat T^{-1} & =& + \hat \tau_z
\ea\ee
It is straightforward to prove that, with the above phases, the operators
(\ref{gene}) satisfy the antilinearity and antihermicity requirements
formulated in Eq. (\ref{anti}).

As a  main consequence of the separability of the
pairing interaction there exists
an average gap parameter:
\be
\Delta_{\alpha\beta} ={1\over 2}\displaystyle \sum_{\gamma\delta}
\bar v_{\alpha\beta\gamma\delta} \kappa_{\gamma\delta} =
-g_{\alpha\beta} \left\{ {G\over 2} \displaystyle \sum_{\gamma\delta}
g_{\gamma\delta}^* \kappa_{\gamma\delta} \right\} \equiv -g_{\alpha\beta}
\Delta
\ee
Using the generators (\ref{gene}) we obtain:
\be
\Delta^{T=1}_{\alpha\tau\widetilde{\beta\tau}} =
-\delta_{\alpha_\tau \beta_\tau} \Delta_{\tau\tau}^{T=1}\quad
\mbox{where} \quad
\Delta_{\tau\tau}^{T=1}=G_{\tau\tau}^{T=1}\displaystyle\sum_{\alpha_\tau > 0}
\kappa_{\alpha_\tau\widetilde{\alpha_\tau}}
\ee
\be\label{del1}
\Delta_{\alpha\tau \widetilde{\beta -\tau}}^{T=1} =
        - \langle \alpha_\tau | \beta_{ -\tau}\rangle
            \Delta_{np}^{T=1}
\quad \mbox{where} \quad \Delta_{np}^{T=1} =
{1\over 2}G^{T=1}_{np} \displaystyle\sum_{\alpha_n\beta_p > 0}
\langle \alpha_n |\beta_p\rangle \left\{
\kappa_{\alpha_n\widetilde{\beta_p}} + \kappa_{\beta_p\widetilde{\alpha_n}}
\right\}
\ee
\be\label{del0}\ba{ccc}
&\Delta_{\alpha\tau \beta -\tau}^{T=0} =
          i \tau \langle \alpha_\tau | \beta_{ -\tau}\rangle
            \mbox{{\bf Im}}( \Delta_{np}^{T=0})   \quad \mbox{and}
\quad
\Delta_{\widetilde{\alpha\tau} \widetilde{\beta -\tau}}^{T=0} =
          -i \tau \langle \alpha_\tau | \beta_{ -\tau}\rangle
            \mbox{{\bf Im}}( \Delta_{np}^{T=0})  & \\
&\mbox{where}\quad \Delta_{np}^{T=0}
={i\over 2} G^{T=0}_{np} \displaystyle\sum_{\alpha_n\beta_p > 0}
\langle \alpha_n |\beta_p\rangle \left\{
\mbox{{\bf Im}} (\kappa_{\widetilde{\alpha_n}\widetilde{\beta_p}})
- \mbox{{\bf Im}} (\kappa_{{\alpha_n}{\beta_p}} \right\})  &
\ea\ee
for {\it pp\/}({\it nn\/}) pairing, T=1, ${\bm \alpha}\tilde{\bm \alpha}$
type of {\it np}-pairing and
T=0, ${\bm {\alpha\alpha}}$ of {\it np}-pairing, respectively.

The single-particle potential ${\bm h}$ takes the following form
\be
   h_{\alpha\beta} = e_{\alpha}\delta_{\alpha\beta} -
\omega j^{(x)}_{\alpha\beta} + \Gamma_{\alpha\beta}
\ee
where the single-particle energies, $e_{\alpha}$, are calculated using a
deformed Woods-Saxon potential~\cite{[Cwi87]}. Nuclear rotation is
included using
the cranking approximation~\cite{[Ing54]} and $\Gamma$ originates
from the  contribution of the pairing interaction
to the single-particle channel.
For {\it pp\/}({\it nn\/}) pairing we have:
\be
\Gamma^{T=1}_{\alpha_{\tau}\beta_{\tau}} =
 -G_{\tau\tau}^{T=1} \rho_{\widetilde{\beta_\tau}\widetilde{\alpha_\tau}} \quad
\Gamma^{T=1}_{\widetilde{\alpha_{\tau}}\widetilde{\beta_{\tau}}} =
 -G^{T=1}_{\tau\tau} \rho_{{\beta_\tau}{\alpha_\tau}} \quad
\Gamma^{T=1}_{{\alpha_{\tau}}\widetilde{\beta_{\tau}}} =
 iG^{T=1}_{\tau\tau} \mbox{{\bf Im}}
(\rho_{{\beta_\tau}\widetilde{\alpha_\tau}})
\ee
For T=1, ${\bm \alpha}\tilde{\bm \alpha}$  {\it np}-pairing we obtain:
\be
\left\{ \ba{c}
\Gamma^{T=1}_{\alpha_{\tau}\beta_{\tau'}} \\
\Gamma^{T=1}_{\widetilde{\alpha_{\tau}}
\widetilde{\beta_{\tau'}}} \\
\Gamma^{T=1}_{\alpha_{\tau}\widetilde{\beta_{\tau'}}}
\ea \right\} =  \displaystyle  {1 \over 2} G^{T=1}_{np}
\sum_{\gamma_{-\tau}\delta_{-\tau'} >0}
\langle \alpha_\tau | \gamma_{-\tau} \rangle \langle\beta_{\tau'} |
\delta_{-\tau'} \rangle \left\{ \ba{c}
 -\rho_{\widetilde{\delta_{-\tau'}}\widetilde{\gamma_{-\tau}}}  \\
 -\rho_{{\delta_{-\tau'}}{\gamma_{-\tau}}}  \\
 i\mbox{{\bf Im}} (\rho_{{\delta_{-\tau'}}\widetilde{\gamma_{-\tau}}})  \\
\ea \right\}
\ee
Finally, for T=0, ${\bm {\alpha\alpha}}$ mode of {\it np}-pairing we get:
\be
\left\{ \ba{c}
\Gamma^{T=0}_{\alpha_{\tau}\beta_{\tau'}} \\
\Gamma^{T=0}_{\widetilde{\alpha_{\tau}}
\widetilde{\beta_{\tau'}}} \\
\Gamma^{T=0}_{\alpha_{\tau}\widetilde{\beta_{\tau'}}}
\ea \right\} = \displaystyle {1 \over 2} G^{T=0}_{np}
\sum_{\gamma_{-\tau}\delta_{-\tau'} >0}
\langle \alpha_\tau | \gamma_{-\tau} \rangle \langle\beta_{\tau'} |
\delta_{-\tau'} \rangle \left\{ \ba{c}
\rho_{{\delta_{-\tau'}}{\gamma_{-\tau}}}  \\
\rho_{\widetilde{\delta_{-\tau'}}\widetilde{\gamma_{-\tau}}}  \\
-i\mbox{{\bf Im}} (\rho_{\widetilde{\delta_{-\tau'}} {\gamma_{-\tau}}})  \\
\ea \right\}
\ee

\section{The Lipkin-Nogami method}\label{s4}

The atomic nucleus is a  mesoscopic system
and as such never undergoes
sharp phase transition. The fluctuations are always of importance
in the transition region. The classical example is the mean-field prediction
of a
sharp superfluid-to-normal phase transition, induced by fast rotation,
which is always smeared out in nature.
The effect can be accounted for theoretically by restoring
particle number and this motivated us to include approximate
number-projection in our model.

In our letter~\cite{[Sat97]} we have  demonstrated that number
projection results in a mixing of the T=0 and T=1 pairing phases already in
$N=Z$ nuclei. This suggests that the exclusiveness of these modes
discussed in the literature is due to the mean-field approximation.
For example, in the
SO(8) model calculations the exact solutions
allow for mixing of T=0 and T=1 phases while not the
gBCS~\cite{[Eng96],[Eng97]}.
Recently, it was shown by Goodman~\cite{[Goo98],[Goo99]} that the
mean-field calculations with a $G$-matrix interaction
performed in a rich model-space
in fact do allow for coexistence of T=0 and T=1 pairing phases.
The number-projection technique used in our calculations
is known as the Lipkin-Nogami (LN) theory. Below, we will outline
the main modifications necessary for applying the LN method to
the non-separable proton-neutron system.

The LN theory attempts to construct the state $|LN\rangle$ where
both linear and quadratic constraints ($\tau\tau'\in \{ p,n\}$):
\be\label{eq11}
\langle LN | LN \rangle =1, \quad
\langle LN | \Delta\hat N_\tau |LN \rangle =0, \quad
\langle LN | \Delta\hat N_\tau \Delta\hat N_{\tau'}|LN \rangle =0
\ee
are simultaneously fulfilled.
Variation over the Lipkin-Nogami state $|LN\rangle$ is equivalent to
a restricted HFB-type variation for the {\it auxiliary\/}
Routhian:
\be\label{eq12}
\delta \langle LN | \hat H^\omega | LN \rangle \equiv
\delta \langle HFB | \hat {\cal H}^\omega | HFB \rangle
\ee
where\footnote{
Obviously, $\lambda^{(2)}_{np} = \lambda^{(2)}_{pn}$ and the symmetric
form of the auxiliary Routhian (\ref{eq13}) is chosen for convenience.}
\be\label{eq13}
\hat {\cal H}^\omega = \hat H^\omega  -
\displaystyle\sum_{\tau} \lambda^{(1)}_\tau
\Delta\hat N_\tau - \displaystyle\sum_{\tau \tau'} \lambda^{(2)}_{\tau\tau'}
\Delta\hat N_\tau \Delta\hat N_{\tau'}
\ee
The LN method is not a variational approximation. Only the
standard linear constraints for the particle number are taken into account
as Lagrange-type variational constraints. The parameters
$\lambda^{(2)}_{\tau\tau'}$ are kept constant during the variational
procedure and eventually adjusted self-consistently using three
additional subsidiary conditions\footnote{In the following, all averages over
$|HFB\rangle$ state will be denoted
as $\langle\quad\rangle$ for simplicity.}:
\be\label{eq14}
\langle \hat{\cal H}^\omega(\Delta\hat N_\tau\Delta\hat N_{\tau'} - \langle
\Delta\hat N_\tau \Delta\hat N_{\tau'} \rangle )\rangle = 0.
\ee
with $\Delta\hat N_{\tau}\equiv \hat N_{\tau} - N_\tau$.

At the point of self-consistency $\hat{\cal H}^\omega_{20} =0$,
and the equations (\ref{eq14}) can be rewritten as
\be\label{eq15}
\langle \hat {\cal H}^\omega | 4 \rangle \langle 4 |
\Delta\hat N_\tau \Delta\hat N_{\tau'} \rangle = 0
\ee
where $|4 \rangle \langle 4|$
denotes the projection onto the
4-quasiparticle space. The LN conditions (\ref{eq15})
depend therefore only on the two-body residual interaction.
These equations form a system of three linear equations:
\be\label{eq16}
\displaystyle\sum_{\tau''\tau'''} A^{\tau''\tau'''}_{\tau\tau'}
x_{\tau''\tau'''} = B_{\tau\tau'}
\ee
where\footnote{The asymmetry is induced by the symmetric form of the
routhian (\ref{eq13}). Note, however, that the corrections
to the $ \Gamma$ and $\Delta$ potentials due to LN terms comes out
to be symmetric, see eq. (\ref{eq21}).}
\be\label{eq17}
x_{\tau\tau}=\lambda^{(2)}_{\tau\tau} \quad \mbox{while} \quad
x_{pn}=2\lambda^{(2)}_{pn}
\ee
and
\be\label{eq18}
A^{\tau''\tau'''}_{\tau\tau'} = \langle \Delta\hat N_{\tau''}
\Delta\hat N_{\tau'''}
| 4 \rangle \langle 4 | \Delta\hat N_{\tau} \Delta\hat N_{\tau'} \rangle
\ee
\be\label{eq19}
B_{\tau\tau'} = \langle \hat V_{two-body}
| 4 \rangle \langle 4 | \Delta\hat N_{\tau} \Delta\hat N_{\tau'} \rangle
\ee
The $x_{\tau\tau'}$ (or $\lambda^{(2)}$) parameters are therefore
equal to:
\be\label{eq20}
x_{\tau\tau'} = {{Det{\bm A}}^{\tau\tau'} \over {Det {\bm  A}}}
\ee
where  ${\bm A}^{\tau\tau'}$ denotes the matrix obtained from matrix {\bm A}
by replacing the ${\tau\tau'}$-th column of {\bm A} by  the vector {\bm B}.

Obviously, the LN theory is technically an analog of the standard HFB
theory but for the routhian (\ref{eq13}). The resulting
LN equations take therefore the form of HFB equations with the single particle
field and pairing field renormalized as follows:
\be\label{eq21}
h_{\tau\tau'}^{LN} \rightarrow h_{\tau\tau'} + 2\lambda^{(2)}_{\tau\tau'}
\rho_{\tau\tau'} \quad \mbox{and} \quad
\Delta_{\tau\tau'}^{LN} \rightarrow \Delta_{\tau\tau'} -
2\lambda^{(2)}_{\tau\tau'} \kappa_{\tau\tau'}
\ee
The contribution to the total energy/routhian due to the LN corrections
can be written as:
\be\label{eq22}
\delta E_{LN} = - \displaystyle \sum_{\alpha\tau ,\beta\tau'}
\lambda^{(2)}_{\tau\tau'} \{ \rho_{\alpha\tau ,\beta\tau'} (
\delta_{\alpha\tau ,\beta\tau'} - \rho_{\beta\tau' ,\alpha\tau} )
 - \kappa_{\alpha\tau ,\beta\tau'} \kappa^*_{\beta\tau' ,\alpha\tau} \}
\ee

One should always bear in mind that the LN theory provides a sort
of optimal HFB-like wave function being different then the $|LN\rangle$
state. Therefore the expectation values of all physical operators
should be recalculated using the following formula:
\be
\langle LN | \hat Q | LN \rangle \equiv
\langle \hat {\cal Q}  \rangle=
\langle \hat Q  \rangle
- \displaystyle\sum_{\tau \tau'} \lambda^{(2)}_{\tau\tau'} (\hat Q)
\langle \Delta\hat N_\tau \Delta\hat N_{\tau'}\rangle
\ee
with $\lambda^{(2)}_{\tau\tau'} (\hat Q)$
dependent on the operator ${\hat Q}$ in question.
These coefficients can be calculated from the following set of five
equations:
\be
\langle \hat{\cal Q} \Delta\hat N_\tau \rangle = 0,  \quad
\langle \hat{\cal Q}(\Delta\hat N_\tau\Delta\hat N_{\tau'} - \langle
\Delta\hat N_\tau \Delta\hat N_{\tau'} \rangle )\rangle = 0.
\ee

\section{Numerical calculations}\label{s5}

An open question in mean-field calculations with $np$-pairing is related
to the strength of the interaction. Whereas
the strength of the $pp$- and $nn$-seniority  pairing force can be
established by a fit
to the odd-even mass differences, very little is known
about the strength of the
$np$-pairing force. Based on isospin invariance arguments,
it seems
well justified to assume
that at the N$\sim$Z line $G^{T=1}_{pp(nn)}\sim G^{T=1}_{np}$. Therefore,
we will take in the following
$G_{np}^{T=1}=(G^{T=1}_{nn}+G^{T=1}_{pp})/2$ with $G^{T=1}_{nn(pp)}$
calculated using the average gap method of Ref.~\cite{[Mol92]}.
Concerning the strength of the isoscalar $np$-interaction we will
present most of our results as a function of the parameter
$x^{T=0}$ that scales the strength of T=0 $np$-pairing interaction
with respect to $G^{T=1}_{np}$ i.e. $x^{T=0} = G^{T=0}_{np}/G^{T=1}_{np}$.
In some cases, however, we will show examples of solutions for
the hamiltonian with broken isospin symmetry.

The discussion is divided into two parts. In section~\ref{ss51}
we shall concentrate on the solutions for non-rotating cases.
Subsections~\ref{sss511} and~\ref{sss512} briefly recall
major properties of the solutions
of  both unprojected
(lateron called BCS) and number-projected (lateron called BCSLN)
versions of the  model. In subsection~\ref{ss52}, we will
discuss a procedure allowing to estimate locally, i.e.
in relatively narrow range of mass number A, a {\sl 'physical'} value
of the scaling parameter $x^{T=0}_0$.
In subsection~\ref{ss53} we shall present the results of cranked, number
projected calculations demonstrating different aspects of
the interplay between mean-field, pairing forces and nuclear rotation.

\subsection{The frequency zero case}\label{ss51}

\subsubsection{Basic properties of the BCS solutions}\label{sss511}

Let us first briefly recall the main properties of the BCS model solutions
without number-projection~\cite{[Sat97],[Eng97]}.
Let us start with the solutions for self-conjugated $N=Z$ nuclei.
Disregarding the Coulomb interaction and assuming that
$G_{pp}^{T=1}=G_{nn}^{T=1}=G_{np}^{T=1}$ leads to
$\Delta_{pp}^{T=1}=\Delta_{nn}^{T=1}=\Delta_0$.
The solutions fall then
into three classes depending on the value of $x^{T=0}$:
\begin{description}
\item{({\it i\/})} For $x^{T=0}<1$ ($G^{T=0}_{np} < G^{T=1}_{np}$) the
T=1 pairing is energetically  favored over the T=0 pairing. The pairing energy
depends only on  $\Delta^2\equiv 2\Delta_0^2 + (\Delta_{np}^{T=1})^2$
and {\it no} energy is gained by activating the T=1 $np$-pairing.
Note that in our model all
pairing phases can be simultaneously present in contrast to the results
presented in Ref.~\cite{[Pit99]}.
\item{({\it ii\/})} The solution at $x^{T=0}=1$
($G^{T=0}_{np}=G^{T=1}_{np}$) is
highly degenerate. The BCS energy depends only on
$\Delta^2\equiv 2\Delta_0^2 + (\Delta_{np}^{T=1})^2
+ |\Delta_{np}^{T=0}|^2$. Also in this limit, no
energy is gained due to $np$-pairing.
\item{({\it iii\/})} The solution at $x^{T=0}>1$
($G^{T=0}_{np} > G^{T=1}_{np}$)
corresponds to a pure T=0 $np$-pairing phase.
\end{description}
For $N\ne Z$ the main features of the BCS solutions can be characterized
as follows:
\begin{description}
\item{({\it iv\/})} The T=1, T$_z$=0 $np$-pairing never forms a collective
phase.
\item{({\it v\/})} There exists a critical value of the $x^{T=0}$ parameter,
$x^{T=0}_{crit}$. For $x^{T=0}<x^{T=0}_{crit}$
only the T=1,$|$T$_z |$=1 solution exist while for
$x^{T=0}\geq x^{T=0}_{crit}$
the T=1,$|$T$_z |$=1 and T=0 phases are mixed.
\item{({\it vi\/})} The value of $x^{T=0}_{crit}$  sharply increases
with increasing neutron/proton excess
$|\mbox{{\bf T}}_z|$. In consequence
there exists a critical value $|\mbox{{\bf T}}_z^{crit}|$ above which there
is no collective solution for $np$-pairing.
Our calculations show that this value  is small
($|\mbox{{\bf T}}_z^{crit}| \sim 2$), see next subsection.
\end{description}

Many of the above properties
can be understood in terms of the variational approach, inherent
to the (generalized) BCS-formalism. For the case of equal weight
of each pair, the system will always choose only that kind of
pairs that result in the lowest energy of the system, i.e. result in a sharp
phase transition from T=0 to T=1 and vice versa.
An interesting feature of the gBCS solution is the 'redundant' role of
the T=1 $np$-pairing in our model when applied to even-even nuclei.
 It does not mix with the other pairing phases  and
only in $N=Z$ nucleus it coexists  with $|$T$_z|$=1 pairing.
Similar properties were found in Ref.~\cite{[Eng97]}.
In the particular case of an $N$=$Z$ nucleus, the coexistence of
T=1 pairing phases at $x^{T=0}$=1 is very fragile.
After switching on the Coulomb force the T=1
$np$-pairing phase essentially vanishes.
On the other hand, for odd-odd nuclei, one expects the T=1, $|$T$_z|$=0
pairing to play a role.

The gBCS solutions for isosymmetric  $N=Z$ nuclei have been analyzed
analytically in Ref.~\cite{[Goo72]} assuming time-reversal invariance
and spherical-symmetry in isospace i.e.
$\langle {\hat{\mbox{{\bf T}}}}_i\rangle$=0 for $i=x,y,z$.
It has been demonstrated
that coherent, $np$-paired, solutions can be obtained for this case
provided that
$ \rho_{\alpha p, \alpha p} = \rho_{\alpha n, \alpha n}$
and
$\kappa_{\alpha p, \overline{\alpha p}} = -
\kappa_{\alpha n, \overline{\alpha n}}$.
The method of Ref.~\cite{[Goo72]} can be generalized to $N\ne Z$
nucleus by extending
symmetry condition to
$\langle {\hat{\mbox{{\bf T}}}}_i\rangle$=0 for $i=x,y$
and $\langle {\hat{\mbox{{\bf T}}}}_z\rangle$=$(N-Z)/2$.
For $N\ne Z$ nuclei, however, neither
$|\rho_{\alpha p, \alpha p}| = |\rho_{\alpha n, \alpha n}|$
nor
$|\kappa_{\alpha p, \overline{\alpha} p}| =
|\kappa_{\alpha n, \overline{\alpha} n}|$.
It is then rather straightforward to show, using the HFB conditions
${\bm \rho}{\bm \kappa} = {\bm \kappa}{\bm \rho}^*$ and
${\bm \rho}^2-{\bm \rho}={\bm \kappa}{\bm \kappa}^*$, that
the gBCS theory does not allow for $np$ T=1 pairing solutions.
For details, see Appendix A.
This result relates the strong limitations of possible solutions
of the generalized BCS theory to the underlying symmetries. In this
particular example the limitations are due
to the time-reversal symmetry and the symmetries
('deformations') in isospace.

In our model, $\langle {\hat{\mbox{{\bf T}}}}_y \rangle$=0,
due to the particular form of the density matrix~(\ref{eq7}).
We could verify numerically for a
number of examples that the BCS solutions for an isospin invariant
hamiltonian are realized at
$\langle {\hat{\mbox{{\bf T}}}}_x \rangle$=$\langle {\hat{\mbox{{\bf T}}}}_y
\rangle$=0  independent
on $x^{T=0}$.
This is not surprising since our interaction is symmetric in
isospin space.
For an isospin broken hamiltonian
with $G^{T=1}_{np} > G^{T=1}_{pp(nn)}$ and with
an overcritical value of $G^{T=1}_{np}$ the solutions for $N\ne Z$ nuclei
are triaxial in the isospace i.e.
with $\langle {\hat{\mbox{{\bf T}}}}_x \rangle$$\ne$0, see also
discussion in Appendix A.

One may further extend the gBCS
model and include an isospin cranking term, $-\mu \hat T_x$.
Such a model
can  be considered as the lowest-order approximate isospin-projected
gBCS theory. One-dimensional isospin rotations
have been studied in Ref.~\cite{[Che78]} in an exactly soluble model including
T=1 pairing correlations only.
It has been demonstrated that the isospin-cranking solutions rather
poorly approximate the exact solutions but
the particle-number projected isospin-cranking model offers
a very good approximation to the exact solution.
Note, that the isospin-cranking model may have several interesting
analogies to the well-studied cranking approximation for spatial rotations.
Due to the transformation rules of the isospin operators
under time-reversal
(\ref{phas}) which are different than those for ordinary angular momentum
operators, there may also exist basic differences.
Neither the possible analogies nor the differences were
studied so far.
The isospin cranking model in its simplest one-dimensional or more
sophisticated three-dimensional version, including both T=0 and T=1
$np$-pairing correlations, may provide important clues on the role
of $np$-pairing in excited configurations.

\subsubsection{Basic properties of the BCSLN solutions}\label{sss512}

The BCSLN solution is qualitatively similar to BCS.
The major differences can be summarized as follow:
\begin{description}
\item{({\it i\/})}
Both the critical value of $x^{T=0}_{crit}$
as well as the 'physical' value of $x^{T=0}_0$, extracted according to the
prescription of subsection~\ref{ss52} are larger in the BCSLN model
than in BCS.  For self-conjugated $N=Z$ nuclei, however, the ratio of
$x^{T=0}_{crit} / x^{T=0}_{0}$ is very similar in both BCSLN and BCS
models while for $\mbox{{\bf T}}_z\ne 0$ this ratio is smaller in
number-projected theory.
\item{({\it ii\/})}
For self-conjugated $N=Z$ nuclei, the solutions of
BCSLN allow for mixing of the T=1, $|$T$_z|$=1
pairing phase with isoscalar $np$-pairing but
not for the collective isovector $np$-pairing phase.
\end{description}

A shift of the critical value of $x^{T=0}_{crit}$ towards
values that are larger
than unity for $N=Z$ nuclei is due to the asymmetric way the
Lipkin-Nogami corrections modify the different pairing channels.
Whereas the 'normal' constraints on proton and neutron number
affect the normal and abnormal densities for protons and
neutrons, respectively, the
constraint on $\Delta NZ$ involves an additional constrain on
 $\rho_{np}$ ($\kappa_{np}$).
It appears
(numerically) that $\lambda_{pn}^{(2)}$ is negative implying  that
the Lipkin-Nogami correction due to $np$-pairing, $\delta E_{np}^{LN}$,
is positive, i.e. repulsive. Both quantities  have opposite sign
as compared to the analogous quantities for the like-particle channel,
see Fig.~\ref{lipkin}. It means, that the effective gap parameters are
weakened in the $np$- and enhanced in
$nn$- and $pp$- channels resulting in the above mentioned shift,
see Eq.~(\ref{eq21}).
It is interesting
to note, however, that the onset of $np$-pairing suppresses
like-particle correlations, increasing
both $\lambda_{p(n)}^{(2)}$
and the (attractive) energy contribution
$\delta E_{p(n)}^{(LN)}$ in such a way that the total LN correction
$\delta E_{tot}^{(LN)}$ increases with increasing strength of
the isoscalar $np$-interaction. Note also, that the LN corrections
and  $\lambda^{(2)}$ parameters
are strongly peaked for $N=Z$ nuclei and decrease rapidly with
the value of $|N-Z|$ (Fig.~\ref{lipkin}).

The $\lambda^{(2)}$ parameters can be approximately estimated as:
\be\label{est}
\lambda^{(2)}_{p(n)} \approx {1\over 2}{\partial^2 E\over \partial Z^2(N^2)}
\quad \mbox{and} \quad
\lambda^{(2)}_{pn} \approx {1\over 2}{\partial^2 E\over \partial ZN}
\ee
where $E(N,Z)$ is the nuclear binding energy. Using the standard
liquid-drop functional form for
$E(N,Z)$ one can show that the leading order contributions to (\ref{est})
for $N\approx Z$ nuclei:
\be
\lambda^{(2)}_{p(n)} \approx - {1\over 9}{a_S\over A^{4/3}} +
{a_I\over A} + 2{a_W\over A}\delta_{N,Z}
\ee
\be
\lambda^{(2)}_{pn} \approx - {1\over 9}{a_S\over A^{4/3}}
-{a_I\over A} - 2{a_W\over A}\delta_{N,Z}
\ee
arise from the surface energy, symmetry energy
and Wigner term, respectively.
Obviously, these estimates are valid for complete HFBLN calculations
including contributions to the $\lambda^{(2)}$ parameters coming from
the particle-hole channel. Nevertheless, the above estimates clearly
show that ({\it i\/}) $\lambda^{(2)}_{p(n)}$ and $\lambda^{(2)}_{pn}$
have opposite signs and that ({\it ii\/})
an enhancement in both $\lambda^{(2)}_{p(n)}$ and $\lambda^{(2)}_{pn}$
is expected for $N=Z$ nuclei due to the singularity of
the Wigner energy. These features are in nice qualitative agreement with
our calculations, see Fig.~\ref{lipkin}.
Whether or not this behavior is realistic; reflects deficiencies of the LN
number projection scheme [see discussion in Ref.~\cite{[Dob93]}] or
calls for isospin-projection remains to be
studied.

\subsection{Estimate of the isoscalar $np$-pairing strength}\label{ss52}

 In our Letter~\cite{[Sat97]} we have demonstrated that the T=0 $np$-pairing
field can yield a microscopic explanation of the Wigner energy in even-even
nuclei. It
naturally accounts for the singularity  of nuclear masses in  $N=Z$ nuclei.
The primary mechanism leading to the Wig/-ner cusp can be viewed as
a generalization of the well-known blocking effect with additional neutrons
or protons outside an $N=Z$ core playing a similar role as the
odd particle plays in the standard blocking phenomenon.
This is visualized schematically in Fig.~\ref{block}.
It needs to be stressed that
the generalized blocking is not a specific property of our schematic
model but does apply for a more realistic pairing interactions
as well, see discussion in~\cite{[Eng96],[Rop00]}.
Moreover, while the standard blocking mechanism requires
different trial wave functions for odd and even systems,
the generalized blocking does not.
The same trial wave function is used
for $N=Z$ and $N\ne Z$ nuclei in the case shown in Fig.~\ref{block}.

The Wigner cusp can be obtained
either by an isospin invariant model with $G^{T=0} > G^{T=1}$ or
isospin broken model with $G^{T=1}_{np} > G^{T=1}_{nn(pp)}$
as shown in~\cite{[Civ97]}. We do not see, however, any particular
reason to invoke an isospin-broken model, particularly in the light of
the discussion  in  Refs.~\cite{[Bre90],[Sat97a]}. In these works
both experimental and theoretical
arguments are given that the Wigner energy
originates essentially from T=0 correlations.
>From these studies it is however not clear
whether it is due to $np$-pairing.  In fact,
in Ref.~\cite{[Pov98]}, the $np$-pairing mechanism was excluded.
Note, however, that all these studies were
performed using the nuclear shell-model. The shell-model
Hamiltonian is usually written in the particle-particle representation
but it can be rewritten in the particle-hole
representation using the Pandya transformation~\cite{[Pan56]}.
Therefore, in the shell-model there is no distinct division
into the pairing- and single-particle field. This division is inherent to
mean-field models only. Indeed, the shell-model definition of pairing
in terms of $L=0, S=1, T=1$ and $L=0, S=1, T=0$ as derived from
the $G$-matrix interaction in Ref.~\cite{[Duf96]}
and used in Ref.~\cite{[Pov98]} to analyze the Wigner energy,
is completely inappropriate from the point of
mean-field model calculations, see also discussion in
Ref.~\cite{[Sat97a],[Wys99]}.

Ref.~\cite{[Sat97a]} shows the technique
to extract the strength, $W(A)$, of the Wigner energy [$E_W(A)=W(A)|N-Z|$]
from experimental data.
Using this method and assuming that the Wigner energy is
indeed due to the isoscalar $np$-pairing correlations
allows us to determine a 'physical'
value of the scaling parameter $x^{T=0}_0=G_{np}^{T=0}/G_{np}^{T=1}$
simply by matching experimental and calculated values of $W(A)$.
Note that this
prescription is by no means limited to our schematic pairing interaction.
Under the assumption that the Wigner energy is indeed due to
isoscalar $pn$-pairing the method can be used for any pairing interaction
to establish an overall scaling factor between isovector and
isoscalar part.

Let us stress again at this point, that the 'physical' value
$x^{T=0}_0$ deduced in this subsection refers only to the  frequency zero
calculations because of the missing T=0
${\bm {\alpha}}\bar{\bm {\alpha}}$ component.
For a pairing force, that
contains both ${\bm {\alpha}}\bar{\bm {\alpha}}$ and
${\bm {\alpha}}{\bm {\alpha}}$ components in the T=0 channel,
the same prescription to determine the
strength can be used. Note that also for this case
the same value of $G_{np}^{T=0}$ will be obtained.
Indeed, due to time reversal invariance, the BCS or BCSLN
energy gain caused by the T=0 pairing
will depend only on the modulus $|\vec\Delta^{T=0}|$ of the total
isoscalar gap, $\vec\Delta^{T=0}\equiv (\Delta^{T=0}_{\alpha\alpha},
\Delta^{T=0}_{\alpha\bar\alpha}$), but not on its direction
(see discussion in Section~\ref{ss51}).

Fig.~\ref{wigner} illustrates calculations for $pf$-shell
$A\approx 48$ nuclei. In the calculations we have taken a constant deformation
of $\beta_2 = 0.25$ and included 30 deformed neutron and proton states.
The realistic values of $x^{T=0}_0$ are estimated to
be $\sim$1.13 and $\sim$1.30 for BCS and BCSLN models, respectively, as
shown in the figure. We performed similar
BCSLN calculations for $A\approx 76$ nuclei assuming
$\beta_2 = 0.40$ and taking 40 deformed neutron and proton states.
In this case we have taken $W(A)=47/A$\,MeV~\cite{[Sat97a]} as a reference
value because experimental masses are not available. The calculations yield
$x_0^{T=0}\sim 1.25$  which would imply that  $G^{T=0}$  has a different mass
dependence than $G^{T=1}$.

In this context one should mention, that the deduced 1/$A$ dependence of
$W(A)$ is essentially due to the fast increase of $W(A)$ for very
light $sd$-shell nuclei.
For $sd$-shell nuclei, however, we expect the T=0 pairing to
have a pronounced  $L=0, S=1$ component.
This component becomes strongly quenched
due to the spin-orbit interaction when entering the $pf$-shell.
The properties of T=0 pairing may change strongly
at the border of the $pf$ and $sd$-shells. The 1/$A$ dependence
of $W(A)$ may  therefore even be an artefact. With the present set of data
one cannot extract a reliable $A$ dependence of the Wigner energy based only
on $pf$-shell and heavier nuclei.
No doubt, that data on masses along the $N=Z$ line are urgently
required in order to
resolve this important question.

The procedure to determine the strength is very
sensitive to even small variations in $x^{T=0}$ particularly for
BCS calculations because we are working in the region of the
phase transition. Indeed,
changing $x^{T=0}_{|\mbox{\scriptsize BCS}}$ by $\sim 2\%$
(in $A\approx 48$ nuclei)
affects the Wigner energy by $\sim 200$\,keV, as demonstrated in
Fig.~\ref{wigner}.  Moreover, although
$x^{T=0}_{0|\mbox{\scriptsize BCSLN}} >
x^{T=0}_{0|\mbox{\scriptsize BCS}}$ in $N=Z$,
$A\approx 48$ nuclei, the ratio
$y\equiv x^{T=0}_0/x^{T=0}_{crit}$  appears
to be very similar in both BCS and BCSLN models.
However, for $\mbox{{\bf T}}_z \ne 0$ we get
$y_{\mbox{\scriptsize BCS}} < y_{\mbox{\scriptsize BCSLN}}$ as shown
in Fig.~\ref{gcrit}. In consequence
$|\mbox{{\bf T}}_{z|\mbox{\scriptsize BCSLN}}^{crit}| >
|\mbox{{\bf T}}_{z|\mbox{\scriptsize BCS}}^{crit}|$ as
expected. The difference is still rather small and
$|\mbox{{\bf T}}_z^{crit}| \sim 2$ independently on the model.
Fig.~\ref{mass} compares calculated (BCSLN model) mass excess
$\Delta E = E(x^{T=0})-E(x^{T=0}=1)$
for Cr and Sr isotopes as a function of $\mbox{{\bf T}}_z$. No pronounced
differences are seen between Cr and Sr
isotopes although there is clear tendency for $\Delta E(\mbox{{\bf T}}_z)$
to be more spread in heavier nuclei, see also Fig.~\ref{gcrit}.

The Wigner energy has an additional repulsive component in odd-odd $N=Z$
nuclei. In~\cite{[Sat97a]} it has been
demonstrated that the strength of this component $d(A)_{T=0}\approx W(A)$ i.e.
both components of the Wigner energy have, most likely, the same origin.
This term can be understood within our pairing
theory provided that the odd-spin,
T=0 states in $N=Z$ odd-odd nuclei
are not treated on the same footing like even-even nuclei but
are interpreted as
two-quasiparticle (2$qp$) configurations involving one
broken $np$-pair.
Indeed, treating even-even and odd-odd nuclei on the same
footing within the generalized BCS(LN) approach gives no
odd-odd versus even-even mass staggering, see Fig.~\ref{dterm}.
The blocking of a T=0 {\it np} quasiparticle, on the other side,
would result in an odd-odd versus even-even mass (T=0) staggering
in $N=Z$ nuclei due to the T=0 pairing energy, analogous to the standard
odd-even mass staggering caused by $T=1$ pairing.
Although the T=0 ground state in odd-odd nuclei is of $2qp$ character,
it is still an open problem of how to construct
a proper trail wave function appropriate for odd-odd nuclei.
It seems   that the isospin restoration
(see discussion in Ref.~\cite{[Fra99a]}) may
play a key role here.

\subsection{The cranking calculations}\label{ss53}

As mentioned previously, our
schematic, separable $np$-interaction
does not allow for T=0 ${\bm \alpha}\tilde{\bm \alpha}$
$np$-pairing.
In a condensate, that is dominated by
T=0 ${\bm \alpha}{\bm \alpha}$ pairs, there is
little resistance
for these pairs to align their angular momenta. Note that this
alignment occurs without invoking
any pair-breaking mechanism. The alignment is a smooth function
of the frequency, and no backbend occurs, see Figure~3 in
our letter~\cite{[Sat97]}.

In the low spin regime, this is of course an unphysical behaviour.
The presence
of T=0 ${\bm \alpha}\tilde{{\bm \alpha}}$ pairing
will invoke the pair-breaking mechanism~\cite{[Wys99]}
(low-$J$ $np$-pairs dominate at low-spin~\cite{[Goo72]}).
However, as long as we
do not use any physical measure sensitive to the isospin of the
$np$-pairs the T=0 ${\bm \alpha}\tilde{\bm \alpha}$ $np$-pairing
can be mocked up  by T=1 ${\bm \alpha}\tilde{\bm \alpha}$ $np$-pairing
and vice-versa via a simple readjustment of the
strength (isospin-broken model).
Indeed, both modes
will respond to nuclear rotation in similar way,
i.e. through the pair breaking mechanism which leads to a suppression
of this field at high spins.

In contrast to the ${\bm \alpha}\tilde{\bm \alpha}$ pairing the
${\bm \alpha}{\bm \alpha}$ type pairing will become
enhanced at high rotational frequencies as Coriolis and centrifugal
forces align more and more pairs along the axis of collective
rotation. Indeed, as shown in~\cite{[Mul82],[Sat97]}, the T=0 phase
composed of high-$J$ $np$-pairs survives
to much higher angular momentum.
This is  in accordance to recent shell model Monte-Carlo
calculations~\cite{[Dea97]}.
At very high spins, our schematic T=0 ${\bm \alpha}{\bm \alpha}$ pairing
is probably somewhat too strong since there is no restriction
in the $J$-values of the pairs that participate in the scattering.
However, in accordance to early calculations, we expect to
pick up the main essence of the high spin behaviour.

Fig.~\ref{omcrit} shows an illustrative
example phase-diagram of the critical frequency (the
frequency corresponding to the
onset of $np$-pairing of ${\bm \alpha}{\bm \alpha}$) versus the strength
of this force. The calculations were performed using the BCSLN model
for the $N=Z$ nucleus $^{48}$Cr at constant deformation and for $\hbar\omega
\leq 2$\,MeV. Filled circles represent calculations with
$G_{pp}=G_{nn}=G_{np}^{\alpha\tilde\alpha}$  while
open circles correspond to an isospin symmetry broken Hamiltonian with
$G_{np}^{\alpha\tilde\alpha}
=1.3G_{pp(nn)}$.  Two things are worth to be noticed:
({\it i\/}) $\hbar\omega_{crit}$ is very sensitive to the strength
parameter and ({\it ii\/}) the alignment (in this case the
alignment of $f_{7/2}$
quasiparticles) triggers the onset of T=0 $np$-pairing setting a natural
threshold ($x^{\alpha\alpha}_{t}$) for the process.  Above
this threshold i.e. for $x^{\alpha\alpha} < x^{\alpha\alpha}_{t}$ the
value of $\hbar\omega_{crit}$ increases very sharply. Therefore,
even  the
optimistic scenario would involve the onset of $np$-pairing
(if at all possible) around the crossing frequency i.e. in the
region which is anyhow the most difficult to describe theoretically.
This is in accordance to single-$j$ shell model calculations
\cite{[Wys99],[She99]}.

\subsection{Terminating states}\label{ss54}

In the $pf$-shell, the state of the art $0\hbar\omega$ shell-models
are able to describe nuclear structure
with excellent accuracy up to the maximum spin states (terminating states)
which is built upon the pure $f_{7/2}$ sub-shell, see e.g.
\cite{[Cau95]}.
For example, for the case of $^{48}$Cr,
the state of maximum spin of $I=16^+\hbar$ can be reached within
$\pi f_{7/2}^4\otimes\nu f_{7/2}^4$ configuration.
Particle-hole excitations to $p_{3/2}$ and $f_{5/2}$ sub-shells
allow further to built states of spin up to $I=20^+\hbar$ within
the full $pf$ space. However, already at these spins
the states involving cross-shell particle-hole excitations
are expected to compete.
Unfortunately, with the present-day
shell-model techniques it is not feasible to perform
calculations within the model space needed to describe states above
spin $I=16\hbar$, which in principle should include the whole
$sd$-shell, $pf$-shell and the $g_{9/2}$ sub-shell.

In the standard cranked mean-field model
calculations  the collectivity of the
rotational band in $^{48}$Cr is essentially
exhausted after the simultaneous alignment of $f_{7/2}$ neutrons and protons
\cite{[Cau95]} and one enters the non-collective, unpaired regime
around $I_x=16\hbar$. Building up higher angular momentum states
proceed further in discrete jumps, which are due to crossings
between single-particle down-sloping
and up-sloping routhians at spherical shape. This reoccupation process
will finally lead again to the onset of collectivity.

The presence of T=0
correlations at high angular momenta may strongly alter
the  behaviour of high spin states discussed above.
In our model, above the terminating
$I = 16\hbar$ state we observe
smooth increase of $I_x$ versus $\hbar\omega$ rather than the
step-like process expected in standard scenario, see Fig.~\ref{cr}.
This is due to the
pair scattering from $d_{3/2}$ and $f_{7/2}$ into the
aligned $g_{9/2}$ and $f_{5/2}$ orbits which is {\it entirely\/}
due to T=0 pairing. Partial occupation
of these orbits  triggers
the onset of collectivity after the terminating state~\cite{[Ter98]}.
New experiments targeted to measure the evolution of the
rotational band in $^{48}$Cr beyond $I = 16\hbar$
may therefore shed  light
into the nature of the mean-field T=0 pairing correlations at high spins.
Particularly, the possibility of the T=0 pairing correlations to
affect the electromagnetic transition rates. Indeed,
without T=0 correlations, the wave functions of high-spin states
will have the dominant component of 2p-2h and/or 4p-4h configuration
(due to promotion of a $\pi f_{7/2} \nu f_{7/2}$ pair or
$\pi f_{7/2}^2 \nu f_{7/2}^2$ quartet to higher sub-shells). The
E2-decay from such a configuration into the 0p-0h
($\pi f_{7/2}^4\otimes \nu f_{7/2}^4$) state would then be
strongly hindered.
In contrast, T=0 pairing will cause mixing of these
(multi)particle-(multi)hole configurations (collectivity)
enhancing the E2 transition amplitudes.

\subsection{TRS-calculations at superdeformed shape}\label{ss55}

Nuclear superdeformation (SD) is an extreme case of the spontaneous breaking
of spherical symmetry. In this case our T=0 pairing
interaction, scattering uniformly $np$-pairs, may even be  considered as
quite a realistic approximation provided that the
strength is properly adjusted. The standard
total routhian surface (TRS)
as well as Skyrme-Hartree-Fock calculations
predict  SD bands to appear at relatively low excitation
energy in $N\sim Z$ nuclei of A$\approx$80-90. Predicted deformations
are $\beta_2 \sim 0.55$  for $^{88}$Ru and even as large as
$\beta_2 \sim 0.75$ for $^{88}$Mo, see~\cite{[Bac99],[Wys99a]}.
To illustrate the influence of T=0 interaction on nuclear superdeformation
we performed TRS calculations
for SD $^{88}$Ru. Three variants of the calculations different in the
treatment of pairing have been performed:
({\it i\/}) unpaired single-particle ({\it ii\/}) including
isovector T=1 pairing only, ({\it iii\/}) including both T=0 and T=1
pairing with $x^{T=0}=0.9$, $x^{T=0}=1.1$, and  $x^{T=0}=1.3$.
The dynamical moments of inertia ($J^{(2)}\equiv dI_x/d\omega$)
versus rotational frequency
resulting from this set of calculation are shown in Fig.~\ref{ru}.

As expected, the calculations ({\it i\/}) and ({\it ii\/})
differ strongly at low spins but converge at higher frequencies where
static T=1 pairing correlations are essentially washed out by the
Coriolis force.
The pronounced kink at $\hbar\omega\approx 0.75$\,MeV in calculations
({\it ii\/}) is due to the simultaneous alignment of the $h_{11/2}$ protons
and neutrons.
Choosing the strongly undercritical strength of $G^{T=0}$,
corresponding to $x^{T=0}=0.9$,
results of course in a pair field dominated by T=1 pairs. However,
after the $h_{11/2}$ alignment and the disappearance of static T=1 pairing
correlations, the T=0 correlations start to build up, resulting in
a somewhat larger moment of inertia as compared to the unpaired case.
Clearly, the onset of T=0 pairing is triggered by the alignment.
The $x^{T=0}=1.1$ corresponds
to a slightly undercritical $G^{T=0}$ strength at low $\hbar\omega$. The low
frequency hump in $J^{(2)}$ at $\hbar\omega\approx 0.3$~MeV
is in this case due to the
{\it rotation-induced\/}
phase transition from predominantly T=1 to T=0 pairing~\cite{[Mul82],[Sat97]}.
For larger values of  $x^{T=0}$ the hump moves towards
lower frequencies in accordance with the phase diagram in Fig.~\ref{omcrit}
discussed in Subsect.~\ref{ss53}, and the $h_{11/2}$ alignment is smoothed out.
Moreover, the high-frequency part of $J^{(2)}$ increases
exceeding the 'rigid body' value~\footnote{We use the notation of
'rigid body' value as the one given by calculations without pairing.
Note that this moment
of inertia strongly differs from that of a rigid body, which would be a
constant.}
by up to $\approx$30\%.
Although the increase depends very much on the strength of the
T=0 interaction (and essentially vanishes for $x^{T=0}$$\leq$0.8) the
observation of systematic deviations from the standard T=1 model predictions
can be considered as a rather robust indicator of the importance of the
stretched i.e. ${\bm \alpha}{\bm \alpha}$ type T=0
correlations at high spins.

\section{Summary and conclusions}\label{s6}

Our work presents a number-projected, cranked mean-field formalism
appropriate for the simultaneous
treatment of T=1 and T=0 correlations.
The number-projection technique is a generalization of the Lipkin-Nogami
method.
We use a schematic $np$-pairing
interaction because our interest is to study general properties
of $np$-paired solutions rather than to reproduce properties
of specific nuclei.

Our work shows that the exclusiveness of different pairing phases
is inherent to the BCS method. Already approximate
number-projection leads
to a mixing of the T=0 and T=1 pairing phases. Common for both approaches,
though, is that different pairing phases counteract each other.

It is also demonstrated that neutron or proton excess
block $np$-pairing. This generalized blocking mechanism causes
$np$-pairing to  vanish rapidly when departing
from the $N=Z$ line. The calculations suggest that already at
$|\bm{T}_z|\sim 2$ there is essentially no chance to observe
any collective $np$-pairing. This mechanism serves as a possible
microscopic explanation of the Wigner energy
at the level of mean-field theory.
The even-even versus odd-odd mass staggering
in T=0 states of $N=Z$ nuclei indicates a  $2qp$ structure
(one broken $np$-pair) of the T=0 states in odd-odd $N=Z$ nuclei.
This will require blocked calculations which
are beyond the scope of our present paper.

Our model predicts that the T=0 correlations
do not vanish at high spin, in accordance to previous
calculations. In the high-spins regime stretched T=0 pairing
is predicted to be dominant. The onset of this pairing phase
seems to be triggered by the alignment of high-$j$ quasiparticles
and the related quenching of isovector pairing correlations.
The presence of T=0 pairing at high spins is predicted to
increase the moments
of inertia as well as to affect the evolution of rotational
bands beyond their standard terminating states.
The size of this effect depends critically on the strength of the T=0 force
in the ${\bm \alpha}{\bm \alpha}$ channel.
Below a certain limit,
in our case below $x_{\alpha,\alpha}^{T=0}$$\leq$0.8, the
effect will not be seen.

\bigskip

This research was supported in part by
the U.S. Department of Energy
under Contract Nos. DE-FG02-96ER40963 (University of Tennessee),
DE-FG05-87ER40361 (Joint Institute for Heavy Ion Research),
DE-AC05-96OR22464 with Lockheed Martin Energy Research Corp. (Oak
Ridge National Laboratory), by the Polish Committee for
Scientific Research (KBN) under Contract No.~2~P03B~040~14, and
by the Swedish Institute.

\vfill
\newpage

\appendix

\section{The time-reversal invariance, isospace deformations
and the BCS solutions.}

The time-reversal invariance implies
the following relations for the hermitian density matrix and
antisymmetric pairing tensor elements~\cite{[Goo72]}:
\be\label{eq41}
    \left\{ \ba{c}
\rho_{\alpha\tau ,\alpha\tau} = \rho^*_{\alpha\tau ,\alpha\tau}   \\
\rho_{\alpha\tau ,\alpha\tau'} =
\rho^*_{\overline{\alpha}\tau ,\overline{\alpha}\tau'}              \\
\rho_{\alpha\tau ,\overline{\alpha}\tau} = 0    \\
\rho_{\alpha\tau ,\overline{\alpha} -\tau} =
-\rho^*_{\overline{\alpha}\tau ,\alpha-\tau}
    \ea \right\}
         \quad \quad
    \left\{ \ba{c}
\kappa_{\alpha\tau ,\overline{\alpha}\tau} =
\kappa^*_{\alpha\tau ,\overline{\alpha}\tau}  \\
\kappa_{\alpha\tau ,\alpha-\tau}=
\kappa^*_{\overline{\alpha}\tau ,\overline{\alpha} -\tau}    \\
\kappa_{\alpha\tau ,\alpha\tau} = 0    \\
\kappa_{\alpha\tau ,\overline{\alpha} \pm\tau} =
-\kappa^*_{\overline{\alpha}\tau ,\alpha \pm\tau}
    \ea \right\}.
\ee

The standard BCS constraints
$\langle  \Delta \hat N \rangle = \langle \Delta \hat Z \rangle = 0$
automatically imply that
$\langle {\hat{\mbox{{\bf T}}}}_z \rangle =(N-Z)/2$.
Superimposing further symmetry constraint $\langle {\hat{\mbox{{\bf T}}}}_x
\rangle = \langle {\hat{\mbox{{\bf T}}}}_y \rangle =0 $ together with
time-reversal symmetry induced relations (\ref{eq41}) implies:
\be\label{symm}
\langle {\hat{\mbox{{\bf T}}}}_x \rangle   = \displaystyle\sum_{\alpha > 0 }
\mbox{{\bf Re}}(\rho_{\alpha p, \alpha n}) =
\langle {\hat{\mbox{{\bf T}}}}_y \rangle  = \displaystyle\sum_{\alpha >0}
\mbox{{\bf Im}}(\rho_{\alpha p, \alpha n})=0
\quad \Rightarrow \quad
\rho_{\alpha\tau, \alpha -\tau} = 0 \quad \forall \quad \alpha.
\ee
The HFB solutions further demand that the generalized density matrix:
\be
{\bm \Re} =  \left(  \ba{cc} {\bm \rho}  &  {\bm \kappa} \\
             {\bm \kappa}^\dagger  & {\bm 1 - \bm\rho}^* \ea \right)
\ee
obeys the idempotency condition ${\bm \Re}^2 = {\bm \Re}$ or,
equivalently, satisfies the following relations:
${\bm \rho\bm\kappa} = {\bm \kappa\bm\rho}^*$ and
${\bm \rho}^2 - {\bm \rho} = {\bm \kappa\bm\kappa}^*$.
These relations impose six constraints on the density matrix and pairing
tensor elements which all must be fulfilled simultaneously. They can be
written as:
\be\label{c1}
  \kappa_{\alpha\tau, \alpha -\tau} \rho_{\alpha\tau, \alpha -\tau}
 +\kappa_{\alpha\tau, \overline{\alpha} -\tau}
 \rho_{\alpha\tau, \overline{\alpha} -\tau}
 =0
\ee
\be
  \rho_{\alpha\tau, \alpha -\tau} \mbox{{\bf Re}}
(\kappa_{\alpha\tau, \overline{\alpha} -\tau}) + i\mbox{{\bf Im}}(
\kappa_{\alpha\tau, \alpha -\tau}\rho^*_{\alpha\tau, \overline{\alpha} -\tau})
=0
\ee
\be\label{c2}
\kappa_{\alpha\tau, \alpha -\tau}(\rho_{\alpha\tau, \alpha\tau}
-\rho_{\alpha -\tau, \alpha -\tau}) =
\rho_{\alpha\tau, \overline{\alpha} -\tau}(\kappa_{\alpha\tau,
\overline{\alpha}\tau}
-\kappa_{\alpha -\tau, \overline{\alpha} -\tau})
\ee
\be\label{c3}
\kappa_{\alpha\tau, \overline{\alpha} -\tau}(\rho_{\alpha\tau, \alpha\tau}
-\rho_{\alpha -\tau, \alpha -\tau}) =
\rho_{\alpha\tau, \alpha -\tau}(\kappa_{\alpha\tau, \overline{\alpha}\tau}
-\kappa_{\alpha -\tau, \overline{\alpha} -\tau})
\ee
\be\label{c4}
\rho_{\alpha\tau, \alpha\tau}^2 +
|\rho_{\alpha \tau, \alpha -\tau}|^2 + |\rho_{\alpha \tau,
\overline{\alpha} -\tau}|^2
-\rho_{\alpha\tau, \alpha\tau} =
-\kappa_{\alpha\tau, \overline{\alpha}\tau}^2
-|\kappa_{\alpha\tau, \alpha -\tau}|^2
-|\kappa_{\alpha\tau, \overline{\alpha} -\tau}|^2
\ee
\be\label{c5}
\rho_{\alpha\tau, \alpha -\tau}(1-\rho_{\alpha\tau, \alpha\tau}
-\rho_{\alpha -\tau, \alpha -\tau}) =
\kappa_{\alpha\tau, \overline{\alpha} -\tau}(\kappa_{\alpha\tau,
\overline{\alpha}\tau}
+\kappa_{\alpha -\tau, \overline{\alpha} -\tau})
\ee
\be\label{c6}
\rho_{\alpha\tau, \overline{\alpha} -\tau}(1-\rho_{\alpha\tau, \alpha\tau}
-\rho_{\alpha -\tau, \alpha -\tau}) =
-\kappa_{\alpha\tau, \alpha -\tau}(\kappa_{\alpha\tau, \overline{\alpha}\tau}
+\kappa_{\alpha -\tau, \overline{\alpha} -\tau}).
\ee
For $N= Z$ nuclei, the coherent $np$-paired solution can be obtained when
$\rho_{\alpha p, \alpha p} = \rho_{\alpha n, \alpha n}$
and $\kappa_{\alpha p, \overline{\alpha} p} = -
\kappa_{\alpha n, \overline{\alpha} n}$. This is the solution
found by Goodman and we refer reader to
the Ref.~\cite{[Goo72]} for further details.
For $N\ne Z$ nucleus however, neither
$|\rho_{\alpha p, \alpha p}| = |\rho_{\alpha n, \alpha n}|$
nor
$|\kappa_{\alpha p, \overline{\alpha} p}| =
|\kappa_{\alpha n, \overline{\alpha} n}|$.
Using the relations (\ref{c1})-(\ref{c6})
it is straightforward to show that
the additional symmetry (axiality in isospace) (\ref{symm})
rules out $np$-pairing
of ${\bm \alpha} \bar{\bm \alpha}$ type. Indeed, Eqs. (\ref{c3}) and
(\ref{c5}) give
instantly $\kappa_{\alpha \tau \overline{\alpha} -\tau} = 0$.

\vfill
\newpage


\vfill
\newpage

\begin{figure}[t]
\caption{
Value of the {\bf a)}
Lipkin-Nogami parameters $\lambda^{(2)}$  and  {\bf b)}
contributions to the total energy $\delta E_{(LN)}$, calculated for
the $\mbox{{\bf T}}_z = 0$
nucleus $^{48}$Cr and $\mbox{{\bf T}}_z = 1$ nucleus
$^{50}$Cr. The total correction to the binding energy (filled triangles)
and the constituents due to proton/neutron pairing ($\Diamond$) and
$np$-pairing ($\circ$) are shown separately. Note, that the contribution
due to $np$-pairing is effectively repulsive.
}
\label{lipkin}
\end{figure}

\begin{figure}[t]
\caption{Schematic representation of
the generalized blocking mechanism
due to the neutron excess. Shaded area indicates the
single-particle levels which
do not contribute to $np$-pair scattering.
}
\label{block}
\end{figure}

\begin{figure}[t]
\caption{Experimental ($\bullet$) and calculated
strength $W(A)$ of the Wigner energy for $pf$-shell nuclei.
The curves marked as ($\circ$) correspond to the BCSLN model
calculations with $x^{T=0}= 1$ and 1.30, respectively. The
curves labeled by
($\triangle$) denote the BCS model calculations with
$x^{T=0}= 1$, 1.13 and 1.15, respectively.
}
\label{wigner}
\end{figure}

\begin{figure}[t]
\caption
{Values of $x^{T=0}_{0}/x^{T=0}_{crit}$ versus
$\mbox{{\bf T}}_z$. Open circles present the  BCS model results
($x^{T=0}_{0}\approx 1.13$) while solid circles mark the
BCSLN calculations ($x^{T=0}_{0}\approx 1.30$) in Cr nuclei. The
line denoted by ($\star$) represents BCSLN calculations in Sr nuclei
($x^{T=0}_{0}\approx 1.25$).
See text for further details.
}
\label{gcrit}
\end{figure}

\begin{figure}[t]
\caption{Calculated mass excess $\Delta E = E(x^{T=0})-E(x^{T=0}=1)$ as a
function of $\mbox{{\bf T}}_z$. The calculations have been performed for
Cr (open symbols) and Sr (filled symbols) isotopes and for
$x^{T=0}$ = 1.1 ($\triangle$), 1.2 ($\Diamond$), 1.3($\circ$). The line
denoted by ($\star$) corresponds to
the 'physical' value of $x^{T=0}=1.25$ representative for Sr isotopes.
The strength parameters for like-particle correlations were
calculated using the average gap method. 
In our case it
yields $G_{pp} \approx 0.385(0.250)$\,MeV for Z=24(38), respectively.
The difference in the pairing strength,
$\Delta G = G_{nn} -G_{pp} $ versus
$\mbox{{\bf T}}_z$ is shown
in the insert.
}
\label{mass}
\end{figure}

\begin{figure}[t]
\caption{Experimental ($\bullet$) and calculated ($\circ ,\triangle$)
correction to the Wigner energy, $d(A)_{T=0}$.
Note, that both BCS and BCSLN theories give essentially no
contribution. For further detail see text.
}
\label{dterm}
\end{figure}

\begin{figure}[t]
\caption{Frequency $\hbar\omega_{crit}$ for the onset
of the T=0 $\alpha \alpha$
$np$-pairing as a function of the strength
$x^{T=0}$ of this force.
Calculations were performed for $^{48}$Cr
at constant deformation $\beta_2 = 0.25$ using the BCSLN model. The
line marked
by ($\bullet$) shows the calculations with an isospin-symmetric hamiltonian
i.e. for
$G^{T=1}_{pp}=G^{T=1}_{nn}=G^{T=1}_{np}$ while ($\circ$) denotes
the calculations for an isospin-broken hamiltonian with
$G_{np}^{T=1}=1.3G_{pp(nn)}$.
The vertical arrow marks the position of the crossing frequency
of the $f_{7/2}$ quasiparticles calculated without $np$-pairing.}
\label{omcrit}
\end{figure}

\begin{figure}[t]
\caption{Gain in angular momentum, $\delta I_x$, due to the  onset of T=0
$ \alpha \alpha$ $np$-pairing relative to the
predictions of a model
without $np$-pairing.  The quantity $\delta I_x$ was calculated
for $^{48}$Cr ($\beta_2 = 0.25$) at  high frequency, $\hbar\omega = 2$\,MeV
and is shown as a function of the strength parameter $x^{T=0}$.}
\label{cr}
\end{figure}

\begin{figure}[t]
\caption{The dynamical moment of inertia for the SD band in the
$N=Z$ nucleus $^{88}$Ru
as a function of $\hbar\omega$. Unpaired (T=1 paired) calculations are
marked with $\circ$ ($\bullet$), respectively. The calculations
with T=0 pairing are labeled by $\triangle$
($x^{T=0}$=0.9), $\Box$
($x^{T=0}$=1.1), and $\Diamond$ ($x^{T=0}$=1.3).
Note the large moments of inertia for the T=0 paired solutions at high
frequencies.
}
\label{ru}
\end{figure}

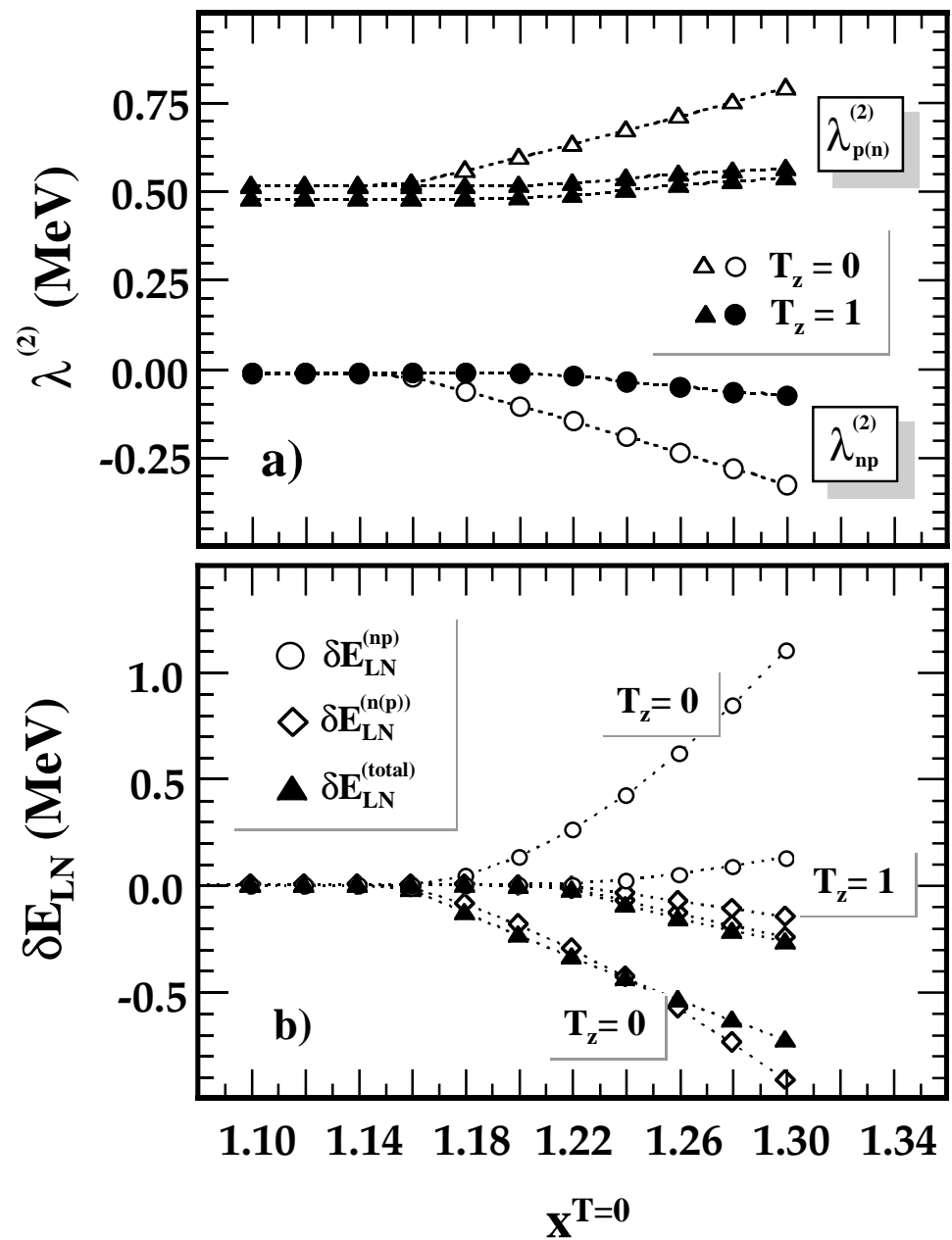
\end{document}